\documentclass[
    11pt,
    letterpaper,
    reprint,
    notitlepage,
    superscriptaddress,
    aps,prx,
]{revtex4-2}

\usepackage{amsmath,amssymb,amsfonts}
\usepackage{physics}
\usepackage{subdepth}
\usepackage{bm}
\renewcommand{\mathbf}{\bm}
\usepackage{dsfont}
\renewcommand{\mathbb}{\mathds}
\usepackage{graphicx}
\usepackage[svgnames,dvipsnames]{xcolor}
\usepackage[colorlinks,linkcolor=red,citecolor=blue,urlcolor=red]{hyperref}
\usepackage[capitalise]{cleveref}
\usepackage[normalem]{ulem}
\usepackage{placeins}
\usepackage[centercolon=true]{mathtools}

\renewcommand{\t}[1]{\mathrm{#1}}

\usepackage{booktabs}
\usepackage{orcidlink}

\newcommand{\LigoMIT}{LIGO Laboratory, Massachusetts Institute of Technology, 185 Albany Street, Cambridge, MA 02139, USA}
\newcommand{\MechMIT}{Department of Mechanical Engineering, Massachusetts Institute of Technology, 
    Cambridge, MA 02139, USA}
\newcommand{\PhysMIT}{Department of Physics, Massachusetts Institute of Technology, Cambridge, MA 02139, USA}
\newcommand{\LLO}{LIGO Livingston Observatory, 19100 Ligo Rd, Livingston, LA 070754, USA}

\begin{document}

\title{Observing and evading quantum back-action on a kilogram-scale oscillator}

\author{Begüm Kabagöz\,\orcidlink{0000-0002-0900-8557}}
\email{begum@mit.edu}
\affiliation{\LigoMIT}
\affiliation{\PhysMIT}
\author{Eric Oelker} 
\affiliation{\LigoMIT}
\affiliation{\PhysMIT}
\author{Dhruva Ganapathy\,\orcidlink{0000-0003-3028-4174}}
\affiliation{Department of Physics, University of California, Berkeley, CA 94720, USA}
\author{Nergis Mavalvala\,\orcidlink{0000-0003-0219-9706}}
\affiliation{\LigoMIT}
\affiliation{\PhysMIT}
\author{Vivishek Sudhir\,\orcidlink{0000-0002-5397-6950}}
\affiliation{\LigoMIT}
\affiliation{\MechMIT}
\author{Vladimir Bossilkov}
\affiliation{\LLO}
\author{Joseph Betzweiser\,\orcidlink{0000-0003-1533-9229}} 
\affiliation{\LLO}
\author{Valery V. Frolov}
\affiliation{\LLO}
\author{Anamaria Effler\,\orcidlink{0000-0001-8242-3944}} 
\affiliation{\LLO}
\author{Adam Mullavey}
\affiliation{\LLO}
% ...

\collaboration{LSC Detector Authors}

\date{\today}

\begin{abstract}
\noindent
Continuous quantum displacement measurements are fundamentally limited by a trade-off between readout imprecision and measurement back-action, constrained by the Heisenberg uncertainty principle. In the Laser Interferometric Gravitational-Wave Observatory (LIGO), these two quantum noise components dominate much of the observation band, making it an excellent testbed. We induce a sub-Hz-linewidth optomechanical mode by trapping the differential motion of the 40-kg mirrors in a band where radiation-pressure back-action dominates the motion. Engineering the quantum state entering the dark port creates correlations between imprecision and back-action that partially cancel their contributions, reducing observed motion near resonance by $\sim 47\%$. A framework resolving the imprecision, back-action, and correlation terms identifies the origin of this suppression. These results demonstrate quantum back-action evasion and quantum reservoir engineering in a macroscopic optomechanical system.

\end{abstract}

\maketitle

The LIGO interferometer measures the differential motion of four $40\,\t{kg}$ mirrors, forming the end-mirrors of a $4\,\t{km}$-long Fabry--P\'erot Michelson interferometer, with a precision of $10^{-20}\, \t{m}/\sqrt{\t{Hz}}$. This precision is necessitated by, and enables, the astrophysical reach of the observatory. Interestingly enough, it also makes LIGO an excellent testbed for macroscopic quantum mechanics. The precision achieved by the interferometer is largely limited by quantum fluctuations
across most of its observation band. 

According to quantum measurement theory \cite{BragVor75,quantum_caves_1981,BraginskyKhalili1992Book}, this exquisite precision must be complemented by back-action force noise arising from the quantum fluctuations of the radiation pressure acting on the mirrors.  
Since both imprecision and back-action noises originate from quantum fluctuations of the optical
field, they can be modified by engineering the quantum state of the light used to perform the measurement, effectively realizing an engineered quantum reservoir for the mirrors.

However, the mirrors, suspended as harmonic oscillators, have a natural frequency ($\sim 0.5\, \t{Hz}$) that is well outside the frequency band over 
which LIGO's precision is quantum noise-limited. Therefore, a broadband rise of quantum noise has previously been observed above the resonance frequency of the suspended mirrors, i.e. in the ``free-mass'' regime, but without separately resolving its constituent sources~\cite{quantum_yu_2020,broadband_ganapathy_2023,squeezing_jia_2024,advanced_capote_2025}. To directly observe the back-action contribution to the differential motion of the mirrors, we define a mechanical mode with resonance frequency where both quantum noises (imprecision and back-action) are significant contributors. We then observe and quantify the quantum back-action from the radiation-pressure quantum fluctuations due to the vacuum and frequency-independent squeezed states of light used for the measurement, and the evasion of that back-action through the use of frequency-dependent squeezing.

\begin{figure*}[hbt!]
	\centering
	{\includegraphics[width=\textwidth]{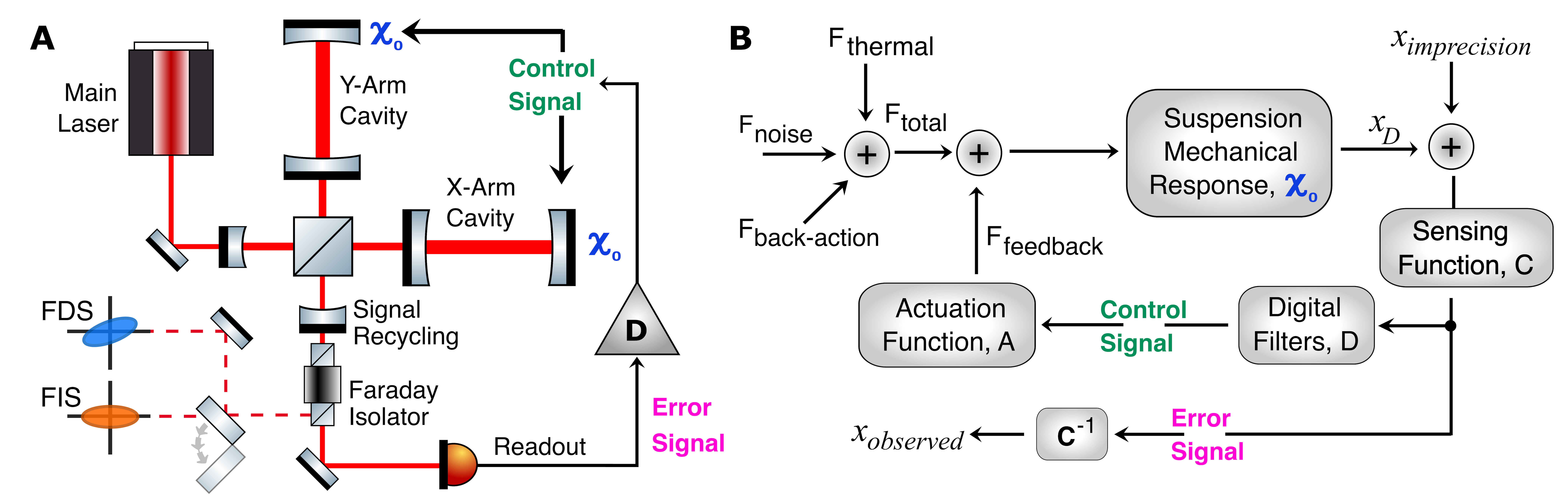}}
    \caption{\textbf{Instrument overview and feedback-loop representation.}
    \textbf{A} Simplified layout of the interferometer and readout chain used to sense and control the feedback-trapped differential arm mode. The main laser is resonant in the arm cavities, and the differential arm displacement is sensed at the antisymmetric port after the signal-recycling cavity and output Faraday isolator. Frequency-independent squeezing (FIS) or frequency-dependent squeezing (FDS) is injected into the readout path to modify the quantum measurement noise and radiation-pressure back-action acting on the mode. The optical readout produces an error signal, which is digitally filtered and converted into a control signal that actuates on the test-mass suspensions.
    \textbf{B} Equivalent linear control diagram for the feedback-trapped mode. Force noise contributions, including thermal force, radiation-pressure back-action, technical force noise are summed to give the total force $F_{\mathrm{total}}$.The total force and the feedback force are acting on the suspension mechanical response $\chi_0$. The resulting displacement $x$ is sensed by the interferometric readout with sensing function $C$, together with measurement imprecision $x_{{imprecision}}$, producing the error signal. The error signal is filtered digitally by $D$ and converted through the actuation path $A$ into the feedback force $F_{\mathrm{feedback}}$, closing the loop. Applying $C^{-1}$ to the error signal gives the calibrated observed displacement $x_{{observed}}$.}
    \label{fig:instrument}
\end{figure*}

\emph{Physics of back-action and its evasion.}
The experiments reported here are performed at the LIGO Livingston Observatory. The differential displacement of the four 40-kg mirrors that form the end-mirrors of a Fabry-Perot-Michelson interferometer (shown in \cref{fig:instrument}A) is transduced into optical phase fluctuations at the anti-symmetric port, which is detected by beating against a local oscillator derived by holding the interferometer at a small static offset~\cite{Fricke2012DCReadout}. The resulting photocurrent fluctuations are a linear record of the differential motion.

The physics of quantum back-action can be illustrated within a simplified lumped-element description, wherein the measured differential degree of freedom of the mirrors can be thought of as a harmonic oscillator whose displacement $x_D(\Omega) = \chi(\Omega)F_\t{tot}(\Omega)$ can be associated with an effective susceptibility $\chi(\Omega)\approx [m(\Omega_\t{m}^2 -\Omega^2 + i \Omega_m^2/Q_m)]^{-1}$ characterized by an effective mass $m \approx 10\, \t{kg}$, resonant frequency $\Omega_\t{m}\approx 0.5\,\t{Hz}$,
and mechanical quality factor $Q\gtrsim 10^8$. In the nominal observatory configuration, the resonant frequency  is much smaller than the gravitational-wave observation frequencies, so that $\chi(\Omega\gg \Omega_\t{m}) \approx -1/(m\Omega^2)$, approximates a free mass. The interferometer senses this motion
with a displacement noise $S_{xx}(2\pi\cdot 150\, \t{Hz}) \approx (2\cdot10^{-20}\, \t{m\sqrt{Hz}})^2$, limited by a combination of quantum fluctuations in the phase of 280 kW  light circulating in the arm cavities and thermal noise of the mirror coatings.  This exquisite measurement precision implies,  by the uncertainty principle \cite{BraginskyKhalili1992Book}
\begin{equation}\label{eq:up}
    S_{xx}^\t{imp}S_{FF}^\t{ba}-\abs{S_{xF}^\t{imp,ba}}^2 \geq {\hbar^2},
\end{equation}
that in the absence of quantum correlations ($S_{xF}=0$), with $S_{xx}^{\t{imp}}$ as quantum imprecision noise and $S_{FF}^{\t{ba}}$ as quantum back-action force noise, the motion must be driven by a quantum back-action force $F_\t{ba}$, originating from quantum fluctuations in the radiation pressure.
Inducing correlations can evade this back-action in principle~\cite{measurement_caves_1980,conversion_kimble_2001}.
However, in the nominal observatory configuration, the measurement imprecision is dominated by classical technical noise around the resonance of the oscillator. Nor is the interferometer's quantum noise simple enough to be described by a single-mode lumped-element model. We address both below.

By reshaping the differential-arm control loop, we can trap the differential mode
at a frequency where the contribution of quantum noise is more 
favorable to observe the interplay of quantum imprecision and back-action~\cite{approaching_whittle_2021}. 
\Cref{fig:instrument}B depicts this feedback loop.
The observed displacement is
$x_\t{obs} = \chi_\t{eff}[F_\t{tot}
+ \chi^{-1}x_\t{imp}]$,
where $\chi_\t{eff}^{-1} = \chi^{-1} + \chi_\t{fb}^{-1}$,
is the effective susceptibility defined by the feedback $\chi_\t{fb}$.
Its spectrum 
\begin{equation}\label{eq:sxx}
    S_{xx}^\t{obs} = |\chi_\t{eff}|^2 S_{FF}^\t{tot} + \Bigl|\frac{\chi_\t{eff}}{\chi}\Bigr|^2 S_{xx}^\t{imp} + 2\,\mathrm{Re}\!\left\{\frac{|\chi_\t{eff}|^2}{\chi}S_{xF}^\t{imp,ba}\right\}.
\end{equation}
makes it clear that the observed motion is shaped not only by the positive imprecision 
and back-action terms, but also by their correlation, which can be negative. 
However, they are constrained by the uncertainty principle in \cref{eq:up}, so that reducing imprecision alone necessarily increases back-action unless correlations are introduced. At first glance, introducing the correlation term inflates the minimum values imprecision and back-action can take. The more striking effect is that the correlation term, when negative, can affect cancellation of the positive back-action 
contribution, and therefore a reduction in the observed motion.
This is the qualitative signature of back-action evasion.

\begin{figure*}[t!]
	\centering
	\includegraphics[width=0.95\textwidth]{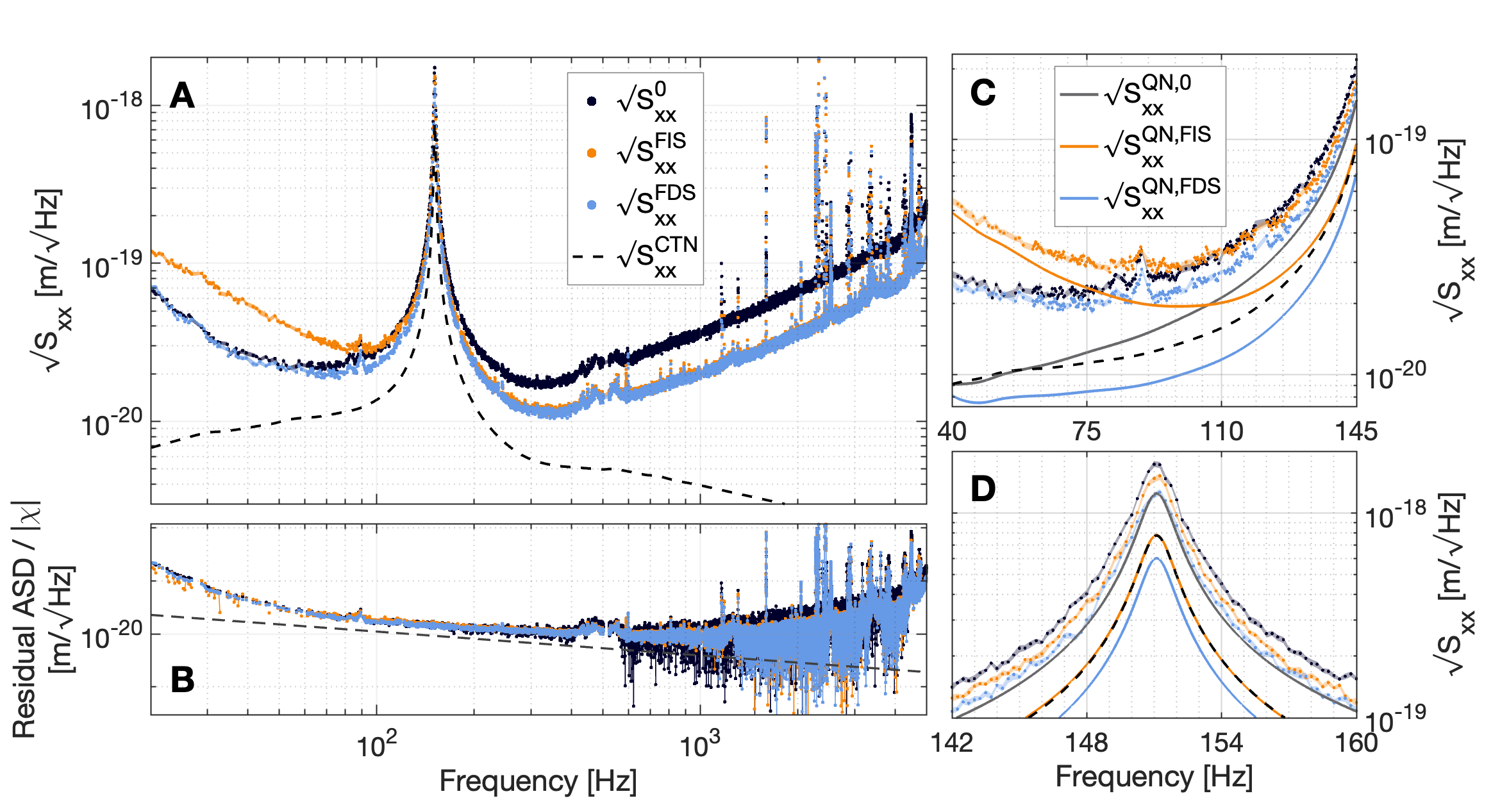}
    \caption{\textbf{Observed displacement spectra and quantum-noise model for the trapped mode.}
    \textbf{A} Observed displacement amplitude spectra for unmodified vacuum (\(\sqrt{S_{\mathrm{xx}}^{0}}\)), frequency-independent squeezing (\(\sqrt{S_{\mathrm{xx}}^{\mathrm{FIS}}}\)), and frequency-dependent squeezing (\(\sqrt{S_{\mathrm{xx}}^{\mathrm{FDS}}}\)), together with the estimated coating thermal noise contribution (\(\sqrt{S_{\mathrm{xx}}^{\mathrm{CTN}}}\)). 
    \textbf{B} Residual classical noise, normalized by the suppression factor \(\mathrm{SF}=\chi_{\mathrm{eff}}/\chi_{0}\), inferred from the measured spectra and the corresponding quantum-noise curves for each input quantum-state configuration. 
    \textbf{C} Zoom of the 40--145 Hz band, showing the inferred quantum-noise model curves for all three configurations together with the coating thermal noise contribution. 
    \textbf{D} Zoom around the trapped-mode resonance. Near resonance, the frequency-dependent squeezing configuration falls below both the unmodified-vacuum and frequency-independent-squeezing cases, consistent with correlation-based suppression of the observed motion. \cite{gwinc} was used to obtain the total quantum noise curves.}
    \label{fig:fds_results}
\end{figure*}

\emph{Observation and evasion of back-action.}
We trap the differential mode of the mirrors at \(151\) Hz with an effective cold-damped quality 
factor of \(\sim 100\). 
These parameters are a result of the unavoidable delays and stability requirements of
the feedback loop used to engineer the trap \cite{approaching_whittle_2021}.
For all ensuing experiments, the configuration of the trap and the mean optical power in the interferometer are held fixed. 

We then compare the motion of the trapped mode for three distinct
quantum states injected at the anti-symmetric port: quantum vacuum, frequency-independent squeezing (FIS)~\cite{quantumenhanced_tse_2019}, and 
frequency-dependent squeezing (FDS)~\cite{broadband_ganapathy_2023,squeezing_jia_2024}.
\Cref{fig:fds_results}A shows the spectrum of the observed motion for these three cases.
The black curve shows the motion of the trapped oscillator when the input state of
the optical field at the anti-symmetric port is ordinary vacuum. 
Injection of frequency-independent phase squeezed vacuum (orange curve) 
lowers the high-frequency imprecision floor, as expected, but increases the 
observed motion at lower frequencies, consistent with enhanced radiation-pressure back-action. Replacing with frequency-dependent squeezing (blue curve) preserves the same 
high-frequency imprecision while suppressing the observed motion in the vicinity and below 
the resonance of the trapped oscillator.
Indeed this behavior is consistent with the fact that FDS in LIGO is engineered to be 
amplitude-squeezed at low frequency and phase-squeezed at high frequency, and that FDS
corresponds to a non-zero correlation $S_{xF}^\t{imp,ba}$. 
The close agreement of the FIS and FDS spectra at high frequency is important: it shows that the resonance-region suppression in the FDS case does not arise from reduction of some 
classical noise or a drift in the level of injected squeezing, 
but in fact from the evasion of back-action due to imprecision-back-action correlations. 

The zoomed panels around the resonance clarifies the nature of back-action further. 
\Cref{fig:fds_results}C shows the 40--145~Hz band, and \Cref{fig:fds_results}D shows 
a narrow window around the resonance. 
In both cases, it is apparent that injection of FDS actually decreases the motion even below
the level of the case of the vacuum. This is consistent with the expectation that 
even in the vacuum case, a fraction of the motion of the trapped oscillator is driven by
back-action~\cite{quantum_buonanno_2001}. Its evasion using FDS verifies its quantum character. 
Around resonance, correlations from the FDS reduce the total displacement-noise power 
by $(47\pm 5)\%$ and $(31\pm 7)\%$ relative to the vacuum and FIS cases respectively.

The solid lines in \cref{fig:fds_results}C,D show the predicted \emph{total} quantum noise 
for each configuration \cite{gwinc}, while the dashed line is the contribution from the coating thermal noise~\cite{CTN}.
The residual between the quantum noise model and the observed data is shown 
in \cref{fig:fds_results}B.
The consistency of the inferred classical residuals across the three configurations 
supports the validity of the quantum-noise curves. 
Their magnitude and spectral shape are also consistent with independent estimates that 
the classical noise in this band is dominated primarily by coating thermal 
noise~\cite{advanced_capote_2025}.

Collectively, these observations are consistent with the interpretation that correlations
due to the FDS may partially cancel the back-action around the resonance of the trapped oscillator.
However, they do not reveal exactly how much the FDS modifies imprecision, 
back-action, and cross correlation to reduce the total noise. Nor do they reveal how much of the change observed in total noise is actually due to a reduction in the differential motion of the optomechanical mode.

\emph{Imprecision and back-action in a multiport system.}
Partitioning the total observed quantum noise into imprecision and back-action is straightforward in a lossless, single-mode, tuned Fabry-Perot-Michelson interferometer, since both originate only from the quantum vacuum field entering the interferometer via its anti-symmetric port, and simple models can fully describe them \cite{conversion_kimble_2001,quantum_buonanno_2001,scaling_buonanno_2003,quantum_yu_2020}.

In a realistic interferometer such as Advanced LIGO, the measured quantum noise is assembled from many coupled pathways: the squeezed input field, vacuum fields entering through multiple loss ports, and the interferometer dynamics~\cite{ligos_mcculler_2021}. 
An accurate description of the measured output is derived from a full 
input-output relation~\cite{quantum_buonanno_2001,ligos_mcculler_2021}
\begin{equation}
    \hat{\mathbf a}_\t{out}(\Omega) = \mathbf{T}_{x}(\Omega)\,\hat{\mathbf{x}} 
    +\mathbf{T}_{\mathrm{sqz}}(\Omega)\,\hat{\mathbf{a}}_{\mathrm{sqz}}
    + \sum_i \mathbf{T}_i(\Omega)\,\hat{\mathbf{a}}_{i,\t{vac}},
\label{eq:multiport}
\end{equation}
where $\hat{\mathbf a}_\t{out}$ is the field leaking out of the interferometer in its locked state, 
\(\mathbf{\hat{x}}\) denotes the residual closed-loop differential and common-mode displacement, including displacement-equivalent noises,  
\(\hat{\mathbf{a}}_{\mathrm{sqz}}\) is the squeezed field injected into the interferometer, 
and \(\hat{\mathbf{a}}_{i,\t{vac}}\) the vacuum fields entering through the various loss ports;
the matrices $\mathbf{T}$ describe the relevant linear responses. 
The photocurrent emitted when the output field is detected takes the form $\hat{\iota}_\t{out}(\Omega) \propto 
\mathbf{e}_\theta^{\mathsf T} \hat{\mathbf a}_\t{out}(\Omega)$, where $\mathbf{e}_\theta$ selects the 
measured quadrature, and is set by the contrast defect of the interferometer and the intentional DC offset in the differential degree of freedom.
Since the photocurrent is linear in the displacement, it can be expressed as,
$\hat{\iota}_\t{out} \propto \mathbf{e}_\theta^{\mathsf T} 
\mathbf{T}_x [\hat{\mathbf{x}}+\hat{\mathbf{x}}_\t{QN}]$,
where $\hat{\mathbf{x}}_\t{QN}$ represents the displacement-equivalent quantum noise. Our interest is in the
differential-mode component of the displacement, which has the schematic form
$x_\t{QN}(\Omega) = \mathbf{\alpha}(\Omega,P)^{\mathsf T} \hat{\mathbf a}(\Omega)$, where 
$\mathbf{\alpha}(\Omega,P)$ is a optical-power-dependent linear response to the set of total 
optical quantum noises collected into \(\hat{\mathbf a}\).
From such a model, we can extract the imprecision, back-action, and correlation contributions by observing that
$\mathbf{\alpha}$ can be separated into two terms with distinct power scaling:
$\mathbf{\alpha}(\Omega,P)= P^{-1/2} \mathbf{\alpha}_{\mathrm{imp}}(\Omega)
+P^{+1/2}\mathbf{\alpha}_{\mathrm{ba}}(\Omega).$

This immediately implies a \textbf{power-law form} for the displacement-referred quantum noise 
\begin{equation}\label{eq:powerlaw}
    S_{xx}^{\mathrm{QN}}(\Omega;P) = \frac{A(\Omega)}{P} + B(\Omega)\,P +C(\Omega),
\end{equation}
where \(A(\Omega)\), \(B(\Omega)\), and \(C(\Omega)\) are independent of \(P\) for fixed optical configuration. 
The first term is the imprecision, the second is the back-action, and the third is their correlation.
Since the form of this expression only relies on the known power-scaling of the imprecision and back-action
noises, and is referenced to the measured output, it is independent of the internal details of the 
interferometer, and holds for any linear multiport interferometer.

\begin{figure}[t!]
	\centering
	\includegraphics[width=0.47\textwidth]{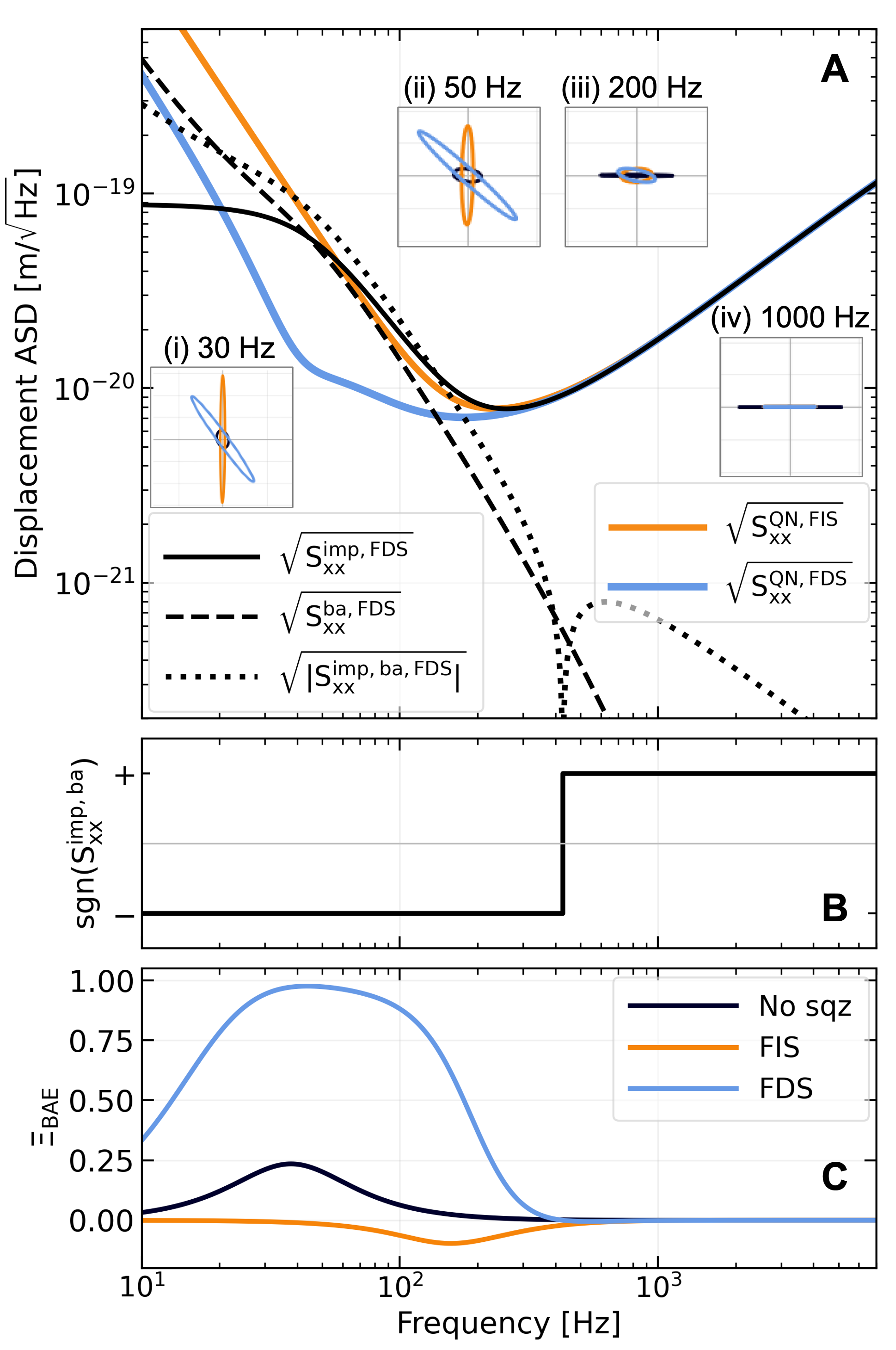}
    \caption{\textbf{Quantum-noise decomposition and correlation-driven back-action evasion.}
    \textbf{A} Decomposition of the inferred displacement quantum noise for the frequency-dependent squeezing (FDS) configuration into the total noise, the effective imprecision contribution \(S_{xx}^\t{imp}\), the effective back-action contribution \(S_{xx}^\t{ba}\), and the magnitude of the correlation term \(\sqrt{|S_{xx}^\t{imp,ba}|}\). The total noise for the frequency-independent squeezing (FIS) configuration is also shown for comparison, highlighting that FIS and FDS provide comparable high-frequency imprecision reduction while differing strongly in the trapped-mode band. Insets show the corresponding effective covariance ellipses in the imprecision--back-action plane (horizontal and vertical axes respectively) at representative frequencies for the unsqueezed, FIS, and FDS configurations; these are not optical quadrature ellipses, but geometric representations of the output-level decomposition. \textbf{B} Sign of the correlation term for the FDS configuration, showing the frequency range over which \(S_{xx}^\t{imp,ba}<0\) and therefore contributes to cancellation of the positive back-action term. \textbf{C} Frequency-resolved back-action-evasion metric \(\Xi_{\mathrm{BAE}}\) for the unsqueezed, FIS, and FDS configurations. The FDS case exhibits substantial correlation-driven cancellation over the same band in which the trapped-mode response is strongest, while the FIS case does not. Full decomposition plots for all three configurations are shown in the Supplementary Material.}
    \label{fig:decomp_bae}
\end{figure}

Operationally, the coefficients \((A,B,C)\) can be extracted from a detailed numerical model of the
interferometer, to result in a physically interpretable decomposition of the displacement quantum noise.
\Cref{fig:decomp_bae}\textbf{A} shows the result of this decomposition for Advanced LIGO:
the FIS configuration (orange) is characterized by significant back-action below 170 Hz, while FDS (blue) exhibits substantial suppression in the same band. The decomposition allows us to partition the quantum noise
and identify that the suppression in the FDS configuration is due to correlation-driven cancellation (black
dotted). Indeed, the correlation term is negative in this band for FDS (\cref{fig:decomp_bae}\textbf{B}). 

To quantify this cancellation, we define a frequency-resolved back-action-evasion metric
\begin{equation}
\Xi_{\mathrm{BAE}}(\Omega)
\equiv
\frac{-S_{xx}^\t{imp,ba}(\Omega)}
{{S_{xx}^\t{imp}}(\Omega)+S_{xx}^\t{ba}(\Omega)},
\label{eq:fbae}
\end{equation}
which estimates the fraction of the positive quantum-noise contribution that is cancelled by correlations 
at each frequency.
\Cref{fig:decomp_bae}\textbf{C} compares \(\Xi_{\mathrm{BAE}}(\Omega)\) for the unsqueezed, FIS, and FDS configurations. The unsqueezed case shows a small effect due to ponderomotive correlations~\cite{quantum_buonanno_2001}. FIS does not improve upon this near the trapped resonance, in fact it produces a small positive correlation that adds to the observed noise rather than cancelling it. Therefore, in the FIS case, the correlations work to augment the total motion of the optomechanical mode, in addition to the already increased radiation pressure back-action contribution to motion, see Supplementary Figure~\ref{fig:qnoise_decomp}. By contrast, FDS is quantitatively different: the engineered frequency-dependent quadrature rotation changes both the magnitude and phase of the imprecision--back-action correlation as well as effective imprecision and back-action so that it yields substantial cancellation in the band where the trapped response is largest. This is the origin of correlation-based back-action evasion, rather than a simple phase rotation of the squeezed quantum state.

\emph{A mechanical oscillator immersed in a squeezed reservoir.} 
The squeezed light effectively modifies the reservoir that the mechanical oscillator interacts 
with \cite{ClarTeufel16}. 
Unlike for a qubit \cite{MurcSiddiqi13c,ToylSiddiqi16}, a squeezed
reservoir modifies only the fluctuations of the oscillator, and not its dissipation.
This is because radiation pressure couples linearly to an optical reservoir quadrature;
squeezing changes the force-noise spectrum, but leaves the force commutator---and
therefore the Kubo response function \cite{Kubo66}---unchanged. 
Indeed a modified fluctuation-dissipation theorem (FDT)
\begin{equation}\label{eq:squeezed_fdt}
    S_{FF}(\Omega) = 2\hbar \left(n_\t{eff}(\Omega) + \tfrac{1}{2}\right) \Im \chi^{-1}(\Omega)
\end{equation}
describes the effect of a squeezed thermal reservoir,
where $n_\t{eff}(\Omega) = N(\Omega) + \Re [M(\Omega) e^{-2i\theta(\Omega)}]$ 
is the effective occupation number given in terms
of the bath occupation $N$, bath correlation $M$, force quadrature angle $\theta$, 
and susceptibility $\chi$
(see Supplementary Information Section \ref{sec:squeezed_bath_fdt}). This form of the FDT implies that the squeezed reservoir can be described by an effective frequency-dependent occupation number, which can be probed through the response of the trapped oscillator.

\begin{figure}[t!]
  \centering
\includegraphics[width=0.4\textwidth]{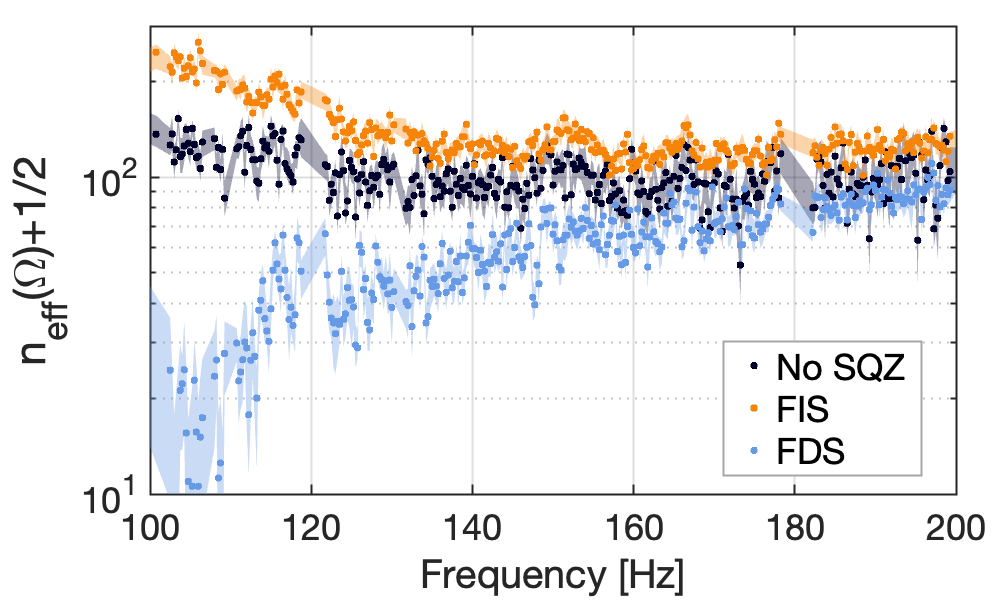}
  \caption{\textbf{Effective phonon occupancy of the reservoir.}
  The effective phonon occupation \(n_\t{eff}(\Omega)+\tfrac{1}{2}\) of the reservoir, appearing
  in the modified FDT \cref{eq:squeezed_fdt}, inferred from the observed motion of the trapped
  oscillator when the reservoir it couples to is unmodified vacuum (black), frequency-independent squeezed (orange), and frequency-dependent squeezed (blue). The error bands are computed from the distribution of segment-wise estimates of the PSD.
  }
  \label{fig:neff_vs_freq}
\end{figure}

\Cref{fig:neff_vs_freq} shows the effective phonon occupation $n_\t{eff}(\Omega)+\tfrac{1}{2}$ 
that appears in the modified FDT. It is inferred from the motion of the trapped oscillator,
estimated by subtracting the imprecision noise from the observed motion, and knowledge of its
susceptibility (see Supplementary).
The depicted quantity partitions the contribution of the reservoir modes at each frequency to
the heating of the oscillator. 
Unlike typical optomechanical systems, 
the notion of a frequency-dependent effective occupation is crucial here, since the 
reservoir is engineered to vary with frequency.
Clearly, a FIS reservoir heats the oscillator at all frequencies to a degree larger than the
vacuum reservoir, while a FDS reservoir has the opposite effect. The latter observation is
consistent with the cancellation of back-action in the observed motion even below the level of the
vacuum case. The effective occupation quantifies precisely how much colder the FDS reservoir
is.

A coarse-grained measure of how hot the oscillator is can be obtained by applying the
equipartition theorem, which relates the total variance of the oscillator's motion to its
effective phonon occupation, a frequency-independent quantity. We do so by integrating the displacement-only noise spectrum, obtained after subtracting the estimated imprecision noise, and referring it to the peak zero-point displacement $S_{xx}^\t{zp}(\Omega_\t{eff}) = \hbar/(2m_\t{eff}\Omega_\t{eff}\Gamma_\t{eff})$, where $\Gamma_\t{eff}\approx 2\pi \cdot 0.74~\t{Hz}$ is the linewidth of the trapped oscillator.
In the limit that the integration bandwidth is much larger than the linewidth, the estimated
occupation converges to \(324\pm 38\) when the vacuum reservoir is unmodified, 
\(423\pm 30\) for frequency-independent squeezed reservoir, and \(198\pm 23\) for 
frequency-dependent squeezed reservoir. 
Residual thermal noise, mostly from the mirror coatings, contributes \(\sim 125\) to the 
occupation, and is what precludes further quantum state engineering of the reservoir.

\emph{Implications and outlook.}
Almost half a century ago, Braginsky identified, and Caves formalized, quantum back-action 
as a fundamental source of noise in gravitational-wave detectors. Yet for decades, it remained
elusive on these instruments, and was instead first observed in tailor-made 
micro-/nano-optomechanical systems. The results reported here close this loop: 
we observe radiation-pressure back-action and demonstrate its correlation-based 
evasion on the very system --- a kilometer-scale interferometer with kilogram-
scale mirrors --- for which the problem was originally posed, using the very mechanism --- frequency-dependent squeezing --- proposed to solve it.

The contribution of this work to the canon of measurement science is threefold. 
First, we demonstrate quantum back-action in the measurement of a 
macroscopic mechanical oscillator, and its evasion by using frequency-dependent squeezing. 
Second, we provide a model-agnostic framework necessary to make such a statement rigorous, 
addressing a problem that only arises at the scale of instrumental complexity exemplified by 
Advanced LIGO. Third, we show precisely the sense in which squeezed light acts as an engineered
reservoir for mechanical motion, and quantify the effective occupation of the reservoir at each
frequency.
In doing so, this work opens the door to controlled preparation of nonclassical motional 
states at the kilogram scale, and is another demonstration that gravitational-wave detectors are not only astrophysical observatories, but also laboratories for macroscopic quantum mechanics.\\ 

\emph{Acknowledgements:} We thank Dong-chel Shin and Hudson Loughlin for comments on the manuscript.
\textbf{Funding:} This material is based upon work supported by LIGO Laboratory which is a major facility fully funded by the National Science Foundation. The authors gratefully acknowledge the support of the US National Science Foundation (NSF) for the construction and operation of the LIGO Laboratory and Advanced LIGO as well as the Science and Technology Facilities Council (STFC) of the United Kingdom, and the Max-Planck-Society (MPS) for support of the construction of Advanced LIGO. Additional support for Advanced LIGO was provided by the Australian Research Council. LIGO was constructed by the California Institute of Technology and Massachusetts Institute of Technology with funding from NSF and
operates under cooperative agreement PHY-2309200. Advanced LIGO was built under award PHY-0823459. The A+ upgrade to Advanced LIGO is supported by US NSF award PHY-1834382 and UK STFC award ST/S00246/1, with additional support from the Australian Research Council. 
VS is funded in part by an NSF CAREER award (PHY–2441238) and  by the Gordon and Betty Moore Foundation (grant GBMF13780). BK is funded in part by NSF PHY-1834382 and by the European Union (ERC, GRAVITES, No. 101071779). 
\textbf{Author contributions:} D.G. and V.S. conceived the project. V.S. and N.M. supervised the project. B.K., V.B., J.B., V.F., A.E., A.M. optimized and configured the LIGO Livingston detector used to collect the data. B.K. performed the data analysis and developed the quantum-noise decomposition model, with E.O. contributing to analysis. V.S. developed the quantum-reservoir-engineering framework. B.K., V.S. and N.M. prepared the manuscript. L.B., E.H., V.F. and P.F. contributed to discussions on the scientific interpretation and experimental design. \textbf{Competing interests:} There are no
competing interests to declare. \textbf{Data and materials availability:}
The data and analysis codes supporting this study are available upon reasonable request. Our best-fitting detector and squeezer model parameters are listed in Table~\ref{tab:qnoise_params}.

\bibliographystyle{apsrev4-2}
\bibliography{ref_v2}

%%%%%%%%%%%%%%%%%%%%%%%%%%%%%%%%%%%%%%%%%%%%%%%%%%%%%%%%%%%%%%%%%%%%%%%
\clearpage

\appendix

\section{Trapped mechanical mode}

The experiment described here is performed at the LIGO Livingston observatory which 
is a 4 km long Michelson-Fabry-Perot interferometer formed by four 40-kg mirrors. 
The interferometer is held at a slight DC offset from the dark fringe by a control loop, 
so that the power leaking out of the antisymmetric port is linearly proportional 
to the differential arm displacement.
In the nominal gravitational-wave (GW) detector configuration, this degree of freedom behaves approximately as a free mass across the GW measurement band. 
In this configuration, the length control loop is designed for stability and robustness, and does not significantly modify the mechanical response of the test masses.

We however reshape the differential-arm control loop to stiffen and trap the differential mode
as a high-$Q$ harmonic oscillator at a frequency around 150 Hz.
Following the quantum description of such feedback controlled system \cite{WilsKipp15,approaching_whittle_2021} gives the observed motion:

\begin{equation}
x_\mathrm{obs}(\Omega)
=
\chi_\mathrm{eff}(\Omega)\,F_\mathrm{ext}(\Omega)
+
\frac{\chi_\mathrm{eff}(\Omega)}{\chi_0(\Omega)}\,x_\mathrm{imp}(\Omega).
\label{eq:xobs_final_trap}
\end{equation}
where $\chi_0$ is the intrinsic mechanical susceptibility of the differential motion to force,
$\chi_\mathrm{eff} = \chi_0/(1+\chi_0 G)$ is the feedback-defined susceptibility with $G$ the
feedback gain, $F_\mathrm{ext}$ is the total force independent of measurement noise, and 
$x_\mathrm{imp}$ is the output-referred measurement imprecision.
With feedback, the imprecision does not merely add at the readout; it is converted into 
an effective force drive through the same dynamics that shapes the trapped-mode motion.

\begin{figure*}[hbt!]
	\centering
	\includegraphics[width=0.95\textwidth]{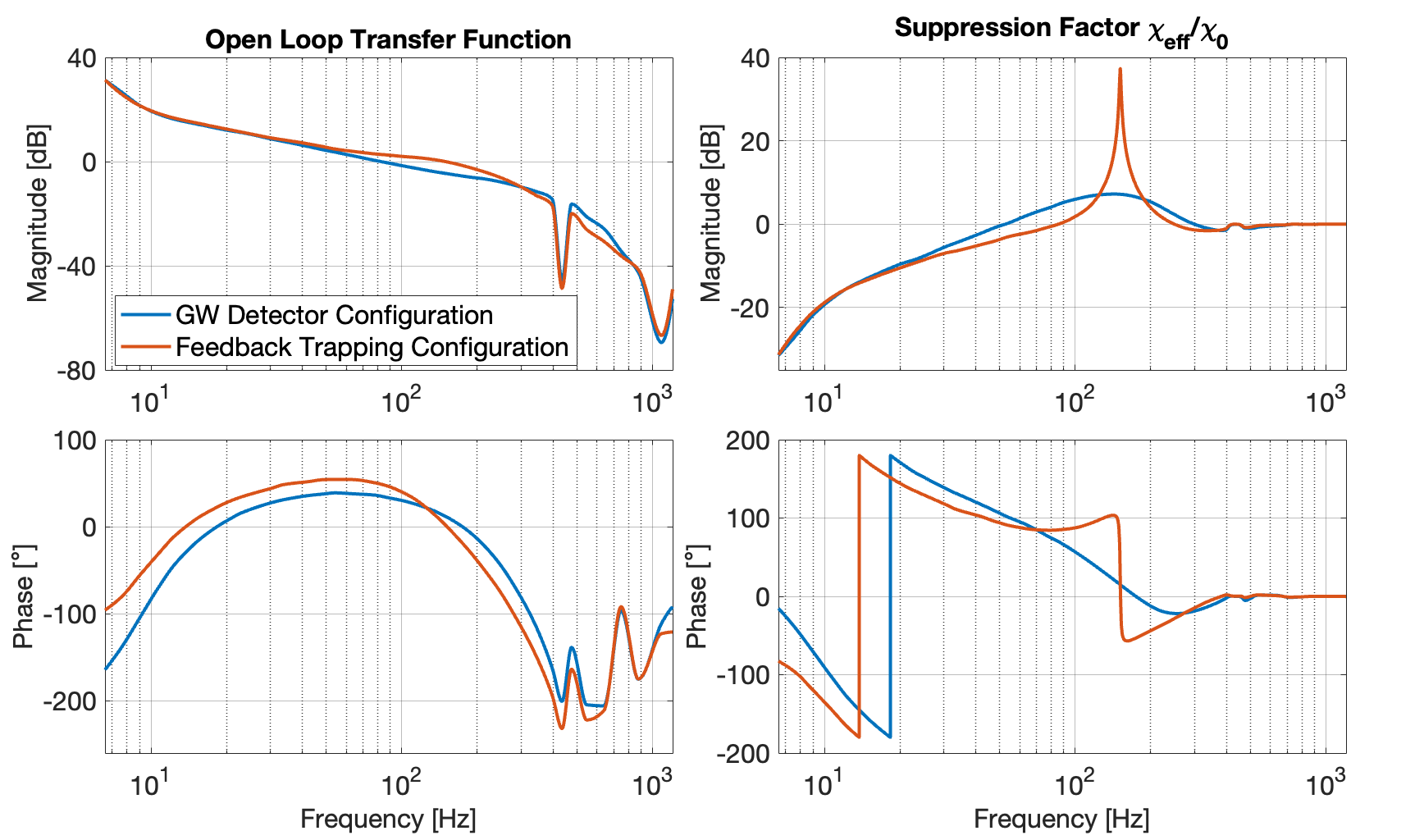}

    \caption{\textbf{Measured DARM open-loop transfer function and feedback-defined susceptibility modification.}
    Left column: magnitude (top) and phase (bottom) of the measured DARM open-loop transfer function (OLTF) for the nominal configuration (no gain peaking) and with the feedback trap engaged. Right column: magnitude (top) and phase (bottom) of the corresponding suppression factor $\chi_\mathrm{eff}(\Omega)/\chi_0(\Omega)$, where $\chi_\mathrm{eff}$ is the effective closed-loop susceptibility inferred from the measured loop response and $\chi_0$ is the nominal free-mass susceptibility. The trapped configuration introduces a narrow-band modification near the trap resonance ($\sim151$~Hz), consistent with an effective harmonic confinement of the differential mode while preserving broadband loop behavior away from resonance.}
	\label{fig:oltf} 
\end{figure*}

The feedback is implemented by digitally shaping the length control loop, resulting in an 
effective response $\chi_\t{eff}$ exhibiting a resonance at $\Omega_\t{eff} \approx 
2\pi\cdot 151$~Hz with a quality factor $Q_\mathrm{eff}\approx 100$, set by delays in the loop.
No other interferometer control parameters are altered when engaging the trap. Optical power, squeezing level, homodyne angle, and alignment settings remain unchanged.

 \Cref{fig:oltf} shows the measured open-loop transfer function 
$L(\Omega) = \chi_0(\Omega) G(\Omega)$ as well as the closed-loop transfer function
$1/(1+L) = \chi_\t{eff}/\chi_0$.
The latter depicts the trapped resonance around 151 Hz. 
Note that $1/(1+L) = \chi_\t{eff}/\chi_0$, which is the factor that converts the estimated displacement (i.e., motion filtered by $\chi_0$) into the observed displacement, as well as
the free-running quantum noise curve (and classical noise contributions) into feedback-trapped observed displacement units. 

\section{Data acquisition and calibration}

The interferometer was maintained in a stable locked state throughout each measurement sequence. Optical power (and therefore the circulating arm power $P_\mathrm{arm}$), alignment settings, homodyne readout angle, and control gains were held fixed unless explicitly modified for the trap configuration.

We first recorded data in the nominal (free-mass) configuration. Time-series of the error signal 
were collected sequentially for the unsqueezed, frequency-independent squeezed (FIS), and frequency-dependent squeezed (FDS) configurations. Switching between squeezing configurations required only modification of the filter cavity operating condition; no interferometer control parameters were changed. The squeezed-beam alignment signals did not change appreciably at the optimal alignment point determined for the FDS configuration, ensuring identical spatial mode matching and optical loss for both FIS and FDS configurations.

Then the feedback-trapped configuration was realized by activating the modified digital filter in the control path. The interferometer was allowed to reach steady state under the new loop conditions before data acquisition began. Data were then recorded sequentially for the unsqueezed, FIS, and FDS configurations under identical trapping parameters.

Following each set of measurements, calibration sweeps were performed to measure the full sensing function and the open-loop transfer function of the DARM control loop. These measurements verify the effective mechanical resonance frequency, linewidth, and loop stability for both the nominal and trapped configurations.

All displacement spectra presented in this work are derived from the differential-arm (DARM) 
calibrated error signal constructed from the two output photodetectors at the 
antisymmetric port~\cite{Cahillane_2017, Sun_2020}.
The calibrated displacement is obtained by dividing the DARM error signal by the 
measured optical sensing function: $x_\t{obs}(\Omega) = x_\t{err}(\Omega)/C(\Omega)$,
where $C(\Omega)$ is the complex sensing response that incorporates the interferometer optical 
gain, the output mode cleaner response, and the digital signal conditioning in the readout path.

The sensing function is measured using swept-sine excitation of the test-mass actuators. A calibrated excitation is applied to the differential mode, and the resulting response in the DARM error signal is recorded to determine  $C(\Omega)$. Calibration measurements were performed within three hours of the start of each data set. Over this interval, variations in the sensing function are below 1\% across the frequency band analyzed here, consistent with independent interferometer characterization studies~\cite{Cahillane_2017,Sun_2020}.

The same sensing function is used to calibrate the unsqueezed, FIS, and FDS data within a given configuration (nominal or trapped). Because the squeezing configurations do not modify the interferometer optical gain or readout electronics, no recalibration is required when switching between squeezing states.

The total systematic uncertainty in the calibrated displacement spectra, including sensing-function measurement uncertainty and known calibration drift, is below 2\% over the frequency range shown in the main text.

\section{Data selection and spectral estimation}

Spectral estimation is performed on the time-series data for each configuration after applying data-quality vetoes and after removing narrow spectral lines unrelated to quantum noise using a known-lines catalog.

\begin{figure*}[hbt!]
	\centering
	\includegraphics[width=0.9\textwidth]{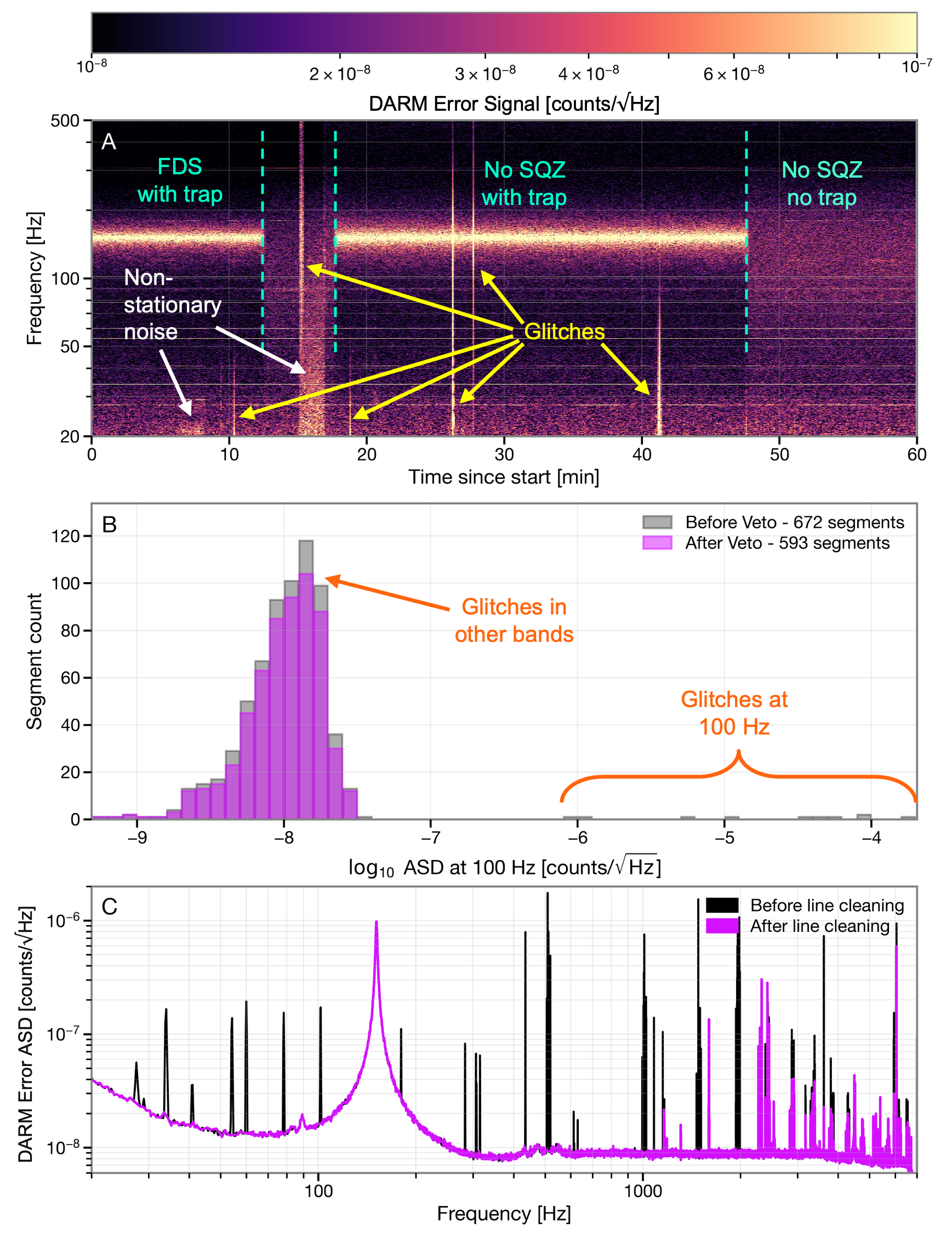}
	\caption{\textbf{A} Spectrogram of an hour long segment during the data campaign, indicating occasional glitches and stretches of time with non-stationary noise. 
    \textbf{B} Histogram for band-limited ASD value around 100 Hz, before (grey) and after (pink) veto. The x axis expresses ASD in logarithmic 10 base.
    \textbf{C} Raw ASD (black) obtained after the veto segments are removed, and line-cleaned ASD (pink), obtained after known lines are filtered from the data.} 
	\label{fig:glitches} 
\end{figure*}

Data quality vetoes exclude known instrumental artifacts and non-stationary behavior that could bias the spectral estimates~\cite{Glanzer2026DetectorCharacterization}.
Specifically, time vetoes were generated from (i) a broadband glitch veto list and (ii) an additional veto list targeting nonstationary intervals identified by excess power in the 100–200 Hz band. For each configuration (nominal vs. feedback-trapped, vacuum vs. FIS vs. FDS injection),  the vetoed time-spans were removed, the remaining ``retained'' spans were merged per configuration and their durations equalized between configurations. This ensures that the final spectra are based on identical effective integration time and are not biased by unequal averaging.

We additionally verified stationarity of the injected squeezing level by monitoring the calibrated displacement spectrum in a high-frequency band around 2 kHz, where radiation-pressure effects are negligible and the spectrum is dominated by imprecision noise. The segment-wise amplitude spectral density in this band was consistent with a stationary squeezing level for both FIS and FDS data sets, with no statistically significant excursions. No additional veto based on squeezing level was required.

Power spectral densities were computed using a Welch-type approach implemented by segmenting the retained time series into overlapping FFTs of duration 32 s with Hann windowing and 60\% overlap. The resulting one-sided PSDs were computed on a fine frequency grid and then restricted to the analysis band (20 Hz to 5 kHz).

Narrow spectral features unrelated to the broadband quantum noise behavior were handled using a known-lines catalog. For each configuration, the catalog was converted into a frequency-bin mask on the fine PSD grid (including single lines and comb families, with configuration-dependent activation windows where specified). Frequency bins within masked regions were excluded from subsequent averaging. To obtain robust broadband spectra for plotting and comparison, the remaining unmasked fine bins were then rebinned using a rolling-window estimator: PSD values were aggregated within 5 Hz-wide frequency windows stepped every 5 Hz, and the median was used as the estimator within each coarse bin. This median-based rebinning suppresses residual sensitivity to narrow artifacts and outliers while preserving the broadband spectral shape.

\FloatBarrier
To quantify statistical uncertainty directly from the data, we computed empirical error bands from the distribution of segment-wise estimates. After rebinning, the PSD (and corresponding ASD) was evaluated for each FFT segment; we report the median ASD and the 16th–84th percentile band across segments as an empirical uncertainty interval. Across the analysis band, these empirical uncertainties are small compared to the systematic calibration uncertainty and do not affect the qualitative comparisons between configurations.

\section{Quantum Noise Model}

The quantum noise contribution to the measured displacement spectra is interpreted using a linear quantum noise model of the interferometer~\cite{ligos_mcculler_2021,gwinc} that includes radiation-pressure coupling, optical losses, and the filter-cavity–induced quadrature rotation. The model propagates vacuum fluctuations entering through the antisymmetric port and distributed loss points through the interferometer response, to compute a total quantum noise term that is made up of the imprecision noise, radiation-pressure force noise, and their correlations at the DARM readout.

Unsqueezed vacuum, frequency-independent squeezing (FIS), and frequency-dependent squeezing (FDS) are modeled within a single framework. The injected state from the dark port (be it unmodified vacuum -- no sqz, or a squeezed state) is parameterized by an effective squeezing level (related to optical loss) and phase. For FIS, the squeezing angle is fixed relative to the interferometer readout quadrature. For FDS, the same squeezed state is reflected from a filter cavity model that imparts a frequency-dependent quadrature rotation determined by the measured cavity detuning and linewidth.

In this manuscript, data from all three configurations are fit simultaneously using a shared parameter set that makes up the effective squeezing level and the effective mechanical parameters of the feedback-defined mode. The only distinction between the FIS and FDS configurations in the model are the frequency-dependent quadrature rotation and mode matching loss imposed by the filter cavity (which is confined to the rotation band). No additional free parameters are introduced to reproduce the behavior near the mechanical resonance.

The model reproduces (i) the identical high-frequency spectra for FIS and FDS, which constrain the effective loss experienced by the squeezed beam, (ii) the increased low-frequency radiation-pressure contribution for FIS, and (iii) the reduction of the resonance-region displacement spectrum in the FDS configuration. The agreement of the models across different configurations confirms that the observed suppression near resonance arises from the engineered quadrature rotation rather than from changes in optical power, loss, or feedback gain.

\begin{figure*}[t!]
  \centering
  \includegraphics[width=\textwidth]{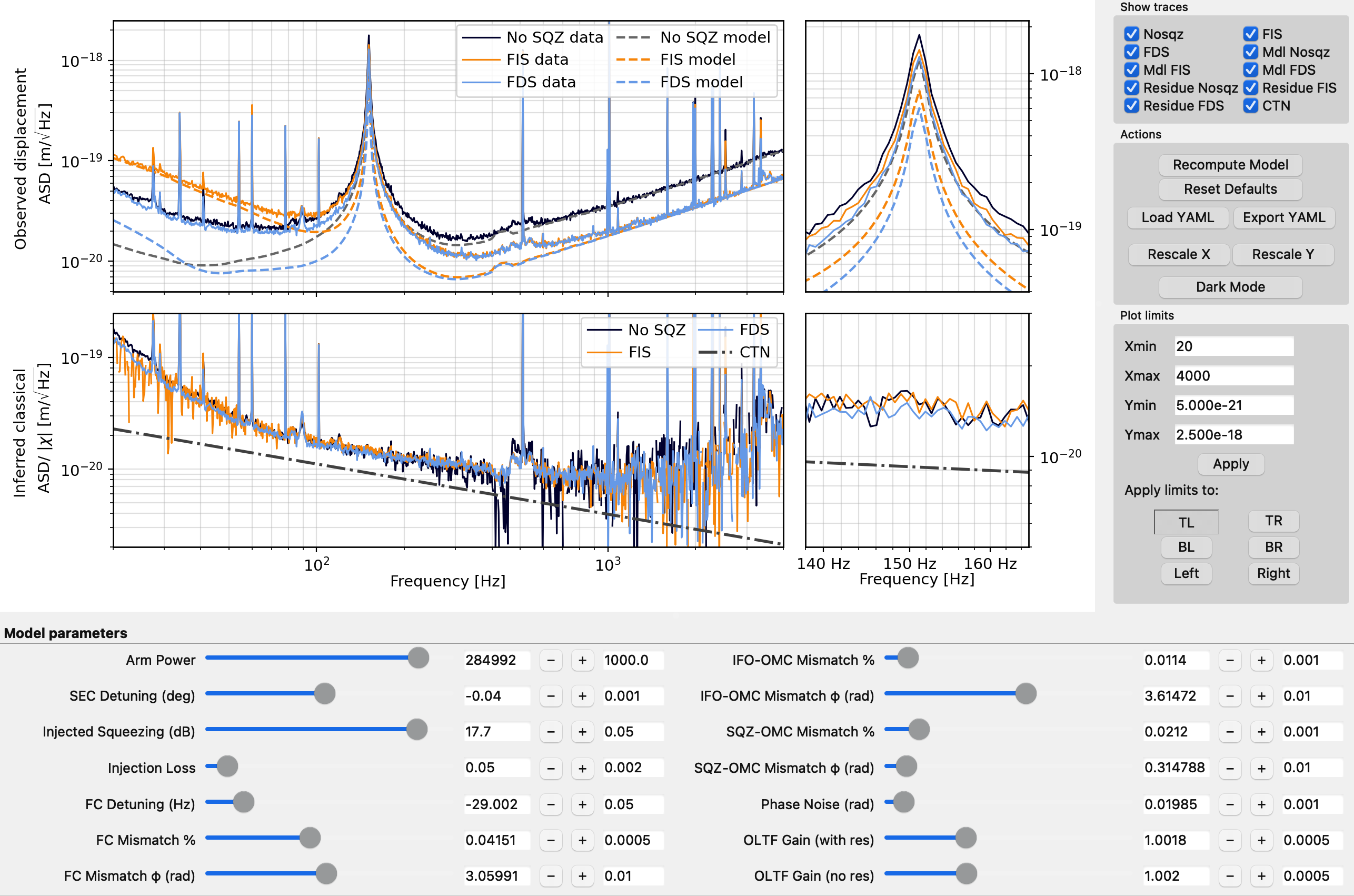}
  \caption{\textbf{Quantum noise fitting interface.}}
  \label{fig:qnoise_fit}
\end{figure*}

The quantum noise model calculation tool pyGWINC ~\cite{gwinc} computes the total quantum noise as a single curve, \textit{without} the capability to separate the individual components imprecision, radiation-pressure and their cross-correlation.

\paragraph*{\textbf{Modification of quantum noise by feedback trapping}}

This section defines how feedback trapping reshapes the quantum-noise contributions to the \emph{observed} displacement
spectrum of the differential arm degree of freedom. The key point is that, in a trapped configuration, sensing
(imprecision) noise is \emph{re-injected as a real force} through the actuation path, so that the total quantum noise
in the measured displacement is filtered by the feedback-defined susceptibility.

We now compute the single-sided PSD of $x_\mathrm{obs}$ from Eq.~(\ref{eq:xobs_final_trap}). Let the force PSD be
$S_{FF}(\Omega)$ for $F_\mathrm{tot}$, the imprecision PSD be $S_{xx}(\Omega)$ for $x_\mathrm{imp}$, and their
cross-spectrum be $S_{xF}(\Omega)$ defined such that
\begin{equation}
\langle x_\mathrm{imp}(\Omega) F_\mathrm{tot}^\ast(\Omega')\rangle
= 2\pi\,\delta(\Omega-\Omega')\,S_{xF}(\Omega).
\end{equation}
Then Eq.~(\ref{eq:xobs_final_trap}) yields

\begin{equation}
 \begin{split}
S_{xx,\mathrm{obs}}(\Omega)
=
|\chi_\mathrm{eff}|^2 S_{FF}(\Omega)
+\Bigl|\frac{\chi_\mathrm{eff}}{\chi_0}\Bigr|^2 S_{xx}(\Omega)\\
+2\,\mathrm{Re}\!\left\{\frac{|\chi_\mathrm{eff}|^{2}}{\chi_0}\,S_{xF}(\Omega)\right\}.
\label{eq:sxx_trapped_decomp}
\end{split}
\end{equation}

Equation~(\ref{eq:sxx_trapped_decomp}) is the trapped-mode analog of the usual shot/RPN/correlation decomposition:
\begin{itemize}
\item \textbf{Back-action (radiation-pressure) pathway:} $|\chi_\mathrm{eff}|^2 S_{FF}$ filters force fluctuations through the
feedback-defined susceptibility. Near the trap resonance, $|\chi_\mathrm{eff}|$ is sharply enhanced, so force-driven motion is amplified.
\item \textbf{Imprecision pathway:} the factor $\bigl|\chi_\mathrm{eff}/\chi_0\bigr|^2$ expresses how sensing noise is converted into observed motion by the feedback loop.
This is the mathematical statement that, in a trapped configuration, imprecision contributes as an \emph{effective actuation drive}.
\item \textbf{Correlation pathway:} the cross term depends on the \emph{phase} of $S_{xF}(\Omega)$ and the complex factor
$\chi_\mathrm{eff}^2/\chi_0$. Frequency-dependent squeezing modifies $S_{xF}(\Omega)$ by rotating optical quadrature correlations,
and the trapped response selects the frequency band where this cross term can coherently cancel part of the back-action contribution.
\end{itemize}

Figure~\ref{fig:qnoise_fit} shows the interface used to fit quantum noise for the trapped mode, with no squeezing, FIS and FDS. The fit fidelity is ensured by monitoring the residual classical noise for each case, while varying parameters shown in the sliders. Table~\ref{tab:qnoise_params} shows the summary of the interferometer, filter cavity and squeezer parameters that are used in this work. 

OpenAI GPT-5.5 Thinking was used to assist with code refinement, debugging, and optimization of the graphical user interface used for fitting. All AI-assisted code was reviewed and validated by the authors. All scientific model choices and parameter interpretation were made by the authors. The LLM was used for code organization and interface iteration and was not used as an independent scientific validation of the analysis.

\begin{table*}[hbt!]
\centering
\caption{Quantum-noise model parameters used here. Parameters that are not listed here are unchanged from Ref.~\cite{squeezing_jia_2024}.}
\label{tab:qnoise_params}
\begin{tabular}{lc@{\hspace{1.2cm}}lc}
\hline\hline
Parameter & Value & Parameter & Value \\
\hline

\multicolumn{2}{l}{\textbf{Interferometer parameters}} &
\multicolumn{2}{l}{\textbf{Squeezing parameters}} \\

Circulating arm power & $285.0~\mathrm{kW}$ &
Generated squeezing & $17.7~\mathrm{dB}$ \\

SEC round-trip detuning phase & $-0.040^\circ$ &
Injection efficiency & $95.0\%$ \\

Readout angle & $-13.9^\circ$ &
Squeezing angle & $13.9^\circ$ \\

IFO--OMC mismatch & $1.14\%$ &
SQZ phase noise (rms) & $19.85~\mathrm{mrad}$ \\

IFO--OMC mismatch phase & $3.615~\mathrm{rad}$ &
SQZ--OMC mismatch & $2.12\%$ \\

\cmidrule(l{0em}r{3.5em}){1-2}
\multicolumn{2}{l}{\textbf{Filter cavity parameters}} &
\raisebox{0.4em}{SQZ--OMC mismatch phase} &
\raisebox{0.4em}{$0.315~\mathrm{rad}$} \\

Detuning & $-29.002~\mathrm{Hz}$ &
\raisebox{0.4em}{SQZ--IFO mismatch (derived)} & \raisebox{0.4em}{$6.23\%$} \\

SQZ--FC mismatch & $4.151\%$ &
& \\

SQZ--FC mismatch phase & $3.060~\mathrm{rad}$ &
& \\

\hline\hline
\label{tab:qnoise_params}
\end{tabular}
\end{table*}

\paragraph*{\textbf{Quantum noise decomposition.}}

\begin{figure*}[t!]
  \centering
  \includegraphics[width=0.9\linewidth]{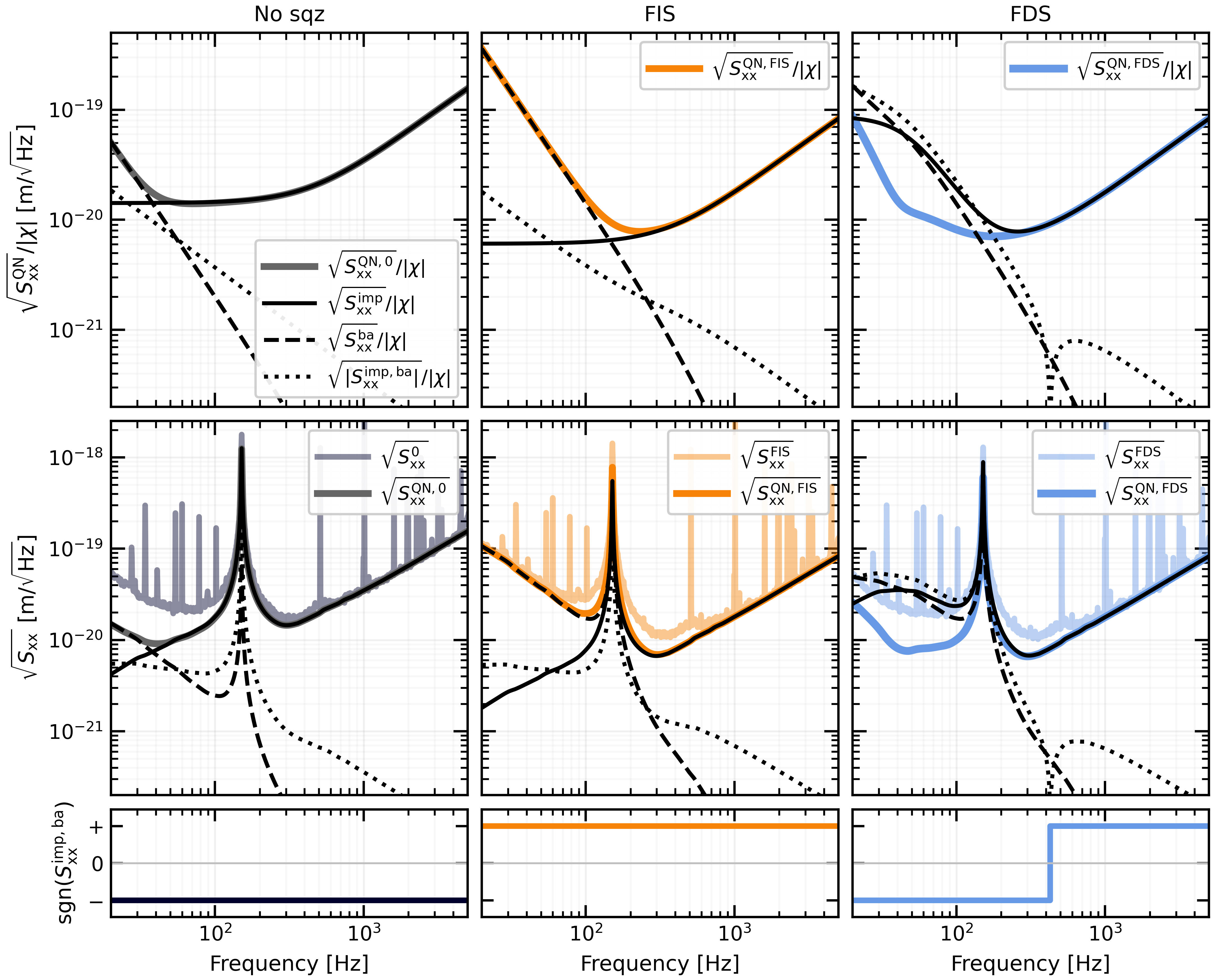}
  \caption{\textbf{Quantum noise decomposed into shot noise, radiation pressure noise and cross-correlation} for estimated displacement (GW detector configuration quantum noise, top), feedback-trapping configuration observed displacement (middle). The bottom panel shows the sign of the cross-correlation component. }
  \label{fig:qnoise_decomp}
\end{figure*}

In the nominal interferometer configuration, the calibrated displacement noise arising from quantum fluctuations can be written as the sum of three contributions,
\begin{equation}
S_{xx}^{\mathrm{QN}}(\Omega)
=
S_{xx}^{\mathrm{shot}}(\Omega)
+
|\chi_0(\Omega)|^2 S_{FF}^{\mathrm{rad}}(\Omega)
+
2\,\mathrm{Re}\!\left\{\chi_0^*(\Omega) S_{xF}(\Omega)\right\},
\end{equation}
where $S_{xx}^{\mathrm{shot}}$ is the measurement imprecision (shot noise), 
$S_{FF}^{\mathrm{rad}}$ is the radiation-pressure force noise, and 
$S_{xF}$ denotes the cross-correlation between imprecision and radiation-pressure fluctuations. 
Here $\chi_0(\Omega)$ is the intrinsic mechanical susceptibility of the differential arm mode.

For a linear interferometric measurement, these quantum noise components exhibit distinct dependence on the circulating arm power $P_\mathrm{arm}$. The imprecision term scales approximately as $S_{xx}^{\mathrm{shot}} \propto 1/{P_\mathrm{arm}},$ while the radiation-pressure force noise scales as $S_{FF}^{\mathrm{rad}} \propto P_\mathrm{arm}.$ 

The cross-correlation term arises from ponderomotive squeezing of the optical field and, in the presence of injected squeezing, from the imposed quadrature rotation; its magnitude is set by the optomechanical coupling and inherits a dispersive frequency dependence.

Because the interferometer operates in the linear regime, this decomposition applies independently at each frequency. The quantum noise model in this manuscript therefore separates the total displacement noise into shot, radiation-pressure, and correlation contributions using their known scaling with $P_\mathrm{arm}$ together with the measured interferometer response. This separation is used in the phonon-number estimation described below.

Figure~\ref{fig:qnoise_decomp} shows decomposition of quantum noise for all three cases, no squeezing, FDS and FIS. The top panel shows the corresponding free-mass quantum noise, the middle panel shows the trapped mode quantum noise and the bottom panel indicated the sign of the cross-correlation term.

The separated components can also be combined to form the back-action–evasion metric defined in Eq.~(\ref{eq:fbae}) of the main text, which quantifies the fraction of the positive quantum-noise contribution cancelled by the correlation term.

\paragraph*{\textbf{Covariance matrix interpretation and ellipse construction.}}
At each analysis frequency $\Omega$, we represent the quantum noise at the output as a two-variable Gaussian process in the basis of
output-referred displacement fluctuations arising from (i) measurement imprecision and (ii) radiation-pressure--driven motion.
We therefore define a (single-sided) covariance matrix for the vector
$\mathbf{x}(\Omega)\equiv\bigl(x_{\mathrm{imp}}(\Omega),\,x_{\mathrm{ba}}(\Omega)\bigr)^{\mathsf{T}}$,

\begin{equation}
\begin{split}
\mathbf{V}_{x}(\Omega)\equiv
\begin{pmatrix}
S_{x_{\mathrm{imp}}x_{\mathrm{imp}}}(\Omega) & S_{x_{\mathrm{imp}}x_{\mathrm{ba}}}(\Omega)\\
S_{x_{\mathrm{ba}}x_{\mathrm{imp}}}(\Omega) & S_{x_{\mathrm{ba}}x_{\mathrm{ba}}}(\Omega)
\end{pmatrix}
\\
=
\begin{pmatrix}
S_{xx}^{\mathrm{shot}}(\Omega) & \tfrac{1}{2}S_{xx}^{\mathrm{imp,ba}}(\Omega)\\
\tfrac{1}{2}S_{xx}^{\mathrm{imp,ba}}(\Omega) & |\chi(\Omega)|^{2}S_{FF}^{\mathrm{rad}}(\Omega)
\end{pmatrix},
\end{split}
\label{eq:cov_matrix}
\end{equation}
where $S_{xx}^{\mathrm{shot}}$ is the output-referred imprecision PSD, $|\chi|^{2}S_{FF}^{\mathrm{rad}}$ is the output-referred
radiation-pressure contribution written in displacement units, and $S_{xx}^{\mathrm{imp,ba}}$ is the correlation contribution to the
measured displacement PSD, such that

\begin{figure*}[t!]
  \centering
  \includegraphics[width=0.8\linewidth]{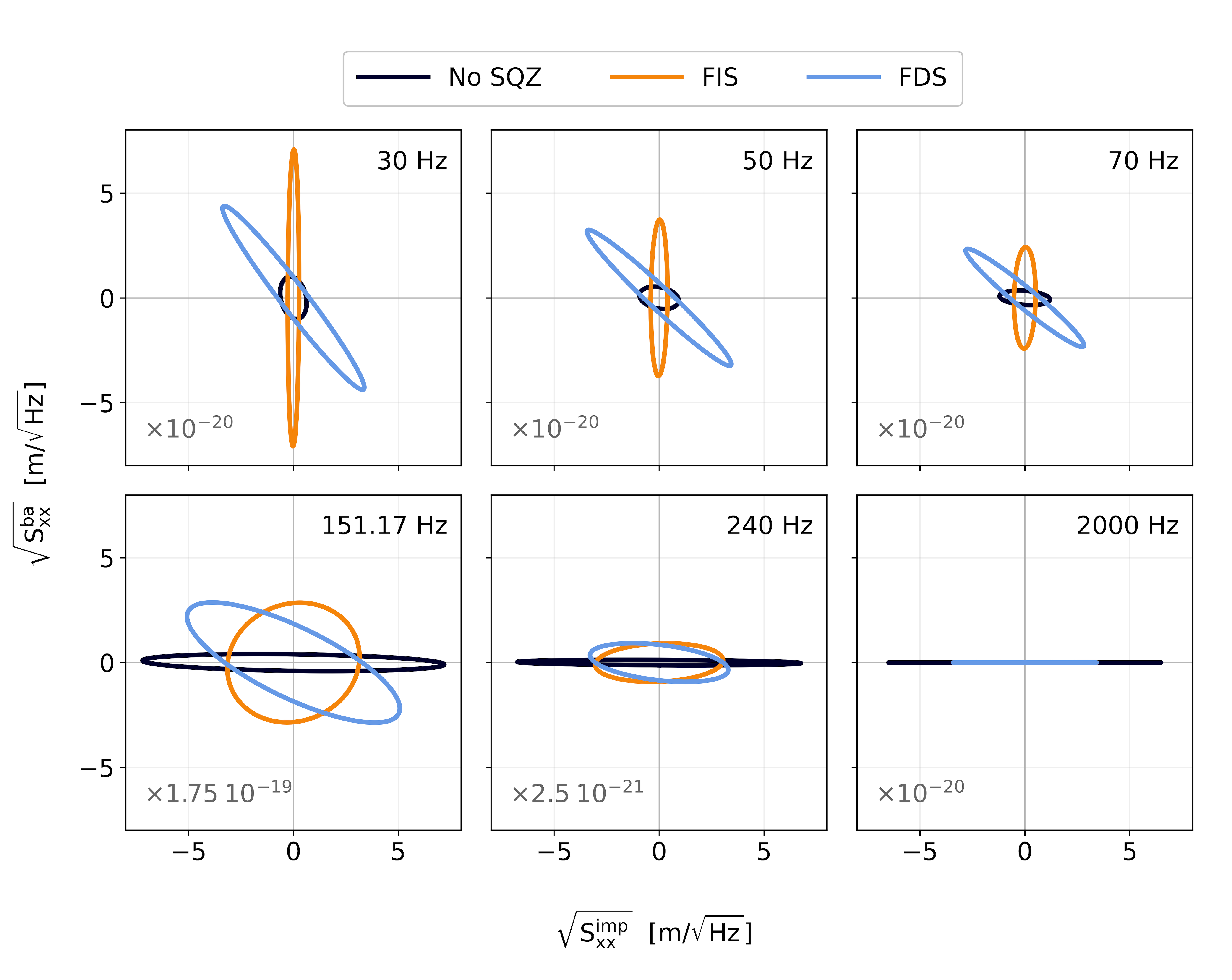}
  \caption{\textbf{Covariance ellipses for chosen frequencies.} The multiplicative factor for both x and y axes for each plot are shown in corresponding bottom left corner.}
  \label{fig:cova}
\end{figure*}

\begin{equation}
S_{xx}^{\mathrm{QN}}(\Omega)=S_{xx}^{\mathrm{shot}}(\Omega)+|\chi(\Omega)|^{2}S_{FF}^{\mathrm{rad}}(\Omega)+S_{xx}^{\mathrm{imp,ba}}(\Omega),
\end{equation}
\begin{equation}
    S_{xx}^{\mathrm{imp,ba}}(\Omega)=2\,S_{x_{\mathrm{imp}}x_{\mathrm{ba}}}(\Omega).
\end{equation}

Figure~\ref{fig:cova} shows the geometric representation of the covariance matrices for no squeezing, FIS and FDS, in the trapped configuration, at selected frequencies. Notice the scale of the 151.17 Hz is different than the other frequencies, due to the resonance peak amplifying all quantum noise components. Regardless, the covariance ellipses at each frequency clearly show the difference between no squeezing, FIS and FDS cases. 

The covariance ellipses shown in this work are constructed directly from
the covariance matrix $\mathbf{V}_x(\Omega)$ at each frequency.
We diagonalize $\mathbf{V}_x(\Omega)$ to obtain its eigenvalues and
eigenvectors. The eigenvectors define the principal axes of the ellipse,
and the square roots of the eigenvalues set the semi-axis lengths.
The plotted curves therefore represent the covariance ellipse in
ASD units (m/$\sqrt{\mathrm{Hz}}$).

\section{Fluctuation--dissipation relation for a squeezed bath}
\label{sec:squeezed_bath_fdt}

The Callen--Welton--Kubo \cite{CallWel51,Kubo66} fluctuation--dissipation theorem (FDT) 
gives the (symmetrized double-sided) spectral density of a force $F$ 
from a thermal bath in terms of the dissipative response of the system to that force:
\begin{equation}\label{eq:fdt_thermal}
    S_{FF}(\Omega) = \hbar \left(2n_\t{th}(\Omega)+1\right) \Im \chi_{FF}(\Omega). 
\end{equation}
Here $n_\t{th}(\Omega) = (e^{\hbar \Omega/k_B T}-1)^{-1}$ is the bath occupation 
at temperature $T$, and 
\begin{equation}
    \chi_{FF}(\Omega) = \frac{i}{\hbar} \int_0^\infty dt\, e^{i\Omega t} \langle [F(t), F(0)] \rangle
\end{equation}
is the linear response to the force. Note that in the main text, we use the displacement
susceptibility $\chi_{xx} = \chi_{FF}^{-1}$.

The FDT assumes that the bath is in a thermal state with respect to the Hamiltonian that 
generates its dynamics. This assumption is violated when the bath is prepared in a squeezed state.
However, as we will now show, the FDT can be generalized to this scenario.

Consider a bath constituted by a continuum of modes $b(\Omega)$ that are bosonic, i.e. 
\begin{equation}\label{eq:bosonicCCR}
    [b(\Omega), b^\dagger(-\Omega')] = 2\pi\delta(\Omega-\Omega').
\end{equation}
Let the bath be prepared in the state, $\rho_\mathrm{sq}=V\rho_\mathrm{th}V^\dagger$,
where $\rho_\t{th}$ is the thermal state, and $V$ is the unitary that squeezes it. 
In particular,
\begin{equation}
    V^\dagger b(\Omega)V = \alpha(\Omega)b(\Omega)+\beta(\Omega)b^\dagger(-\Omega),
\label{eq:sqz_bogoliubov_app}
\end{equation}
such that
\begin{equation}
    |\alpha(\Omega)|^2-|\beta(\Omega)|^2=1,
\end{equation}
to preserve the commutation relations in \cref{eq:bosonicCCR}.
In the squeezed thermal state of the bath,
\begin{equation}
\begin{split}
    \langle b^\dagger(\Omega)b(\Omega')\rangle_\mathrm{sq}
    &=N(\Omega)\cdot 2\pi\delta(\Omega-\Omega'),\\
    \langle b(\Omega)b(\Omega')\rangle_\mathrm{sq}
    &=M(\Omega)\cdot 2\pi\delta(\Omega+\Omega'),
\end{split}
\end{equation}
where
\begin{equation}\label{eq:squeezed_NM}
\begin{split}
    N(\Omega) &=|\alpha(\Omega)|^2n_\mathrm{th}(\Omega)
    +|\beta(\Omega)|^2[n_\mathrm{th}(\Omega)+1] \\
    &=n_\mathrm{th}(\Omega)+|\beta(\Omega)|^2[2n_\mathrm{th}(\Omega)+1],\\
    M(\Omega) &=\alpha(\Omega)\beta(\Omega)[2n_\mathrm{th}(\Omega)+1].
\end{split}
\end{equation}
Clearly, squeezing changes the bath fluctuations by adding both an excess 
occupation $N-n_\mathrm{th}$ and the correlation $M$. 

Let the bath force be proportional to the bath quadrature at angle $\theta$, i.e.
\begin{equation}\label{eq:Ftheta_bath}
    F_\theta(\Omega)=G(\Omega)
    \left[e^{-i\theta}b(\Omega)+e^{i\theta}b^\dagger(-\Omega)\right].
\end{equation}
The fluctuations in it are quantified by its symmetrized spectrum
\begin{align}\label{eq:squeezed_force_noise_app}
    S_{F_\theta F_\theta}(\Omega)
    &=|G(\Omega)|^2\left[2N(\Omega)+1
    +2\mathrm{Re}\{M(\Omega)e^{-2i\theta}\}\right]
\end{align}
Note that the phase of $M$ determines the contribution of the squeezed bath quadrature:
for fixed $N$ and $|M|$, the minimum and maximum of $S_{F_\theta F_\theta}$ are proportional to
$2N+1\pm 2|M|$ respectively. 

By contrast, the dissipative response is determined by the commutator
\begin{equation}
    \chi_{F_\theta F_\theta}(t)=\frac{i}{\hbar}\Theta(t)\langle[F_\theta(t),F_\theta(0)]\rangle .
\end{equation}
For a bath force linear in the bosonic bath modes, as in \cref{eq:Ftheta_bath}, the force
commutator is proportional to the commutator of the bath modes, $[b(\Omega),b^\dagger(\Omega')]$.
But the latter is a scalar constant [see \cref{eq:bosonicCCR}], so that it is unchanged by
the squeezing unitary.
Squeezing therefore does not modify the dissipation. 

To relate $G$ to $\Im\chi_{F_\theta F_\theta}^{-1}$, we use the fact that \cref{eq:squeezed_force_noise_app} must reduce to the thermal FDT in the unsqueezed limit. Setting $\alpha =1, \beta=0$ gives
$N=n_\t{th}$ and $M=0$, and so the unsqueezed force spectrum is
$S_{F_\theta F_\theta}^\t{th} = |G|^2[2n_\t{th}+1]$. 
Comparing this to the thermal FDT expression in \cref{eq:fdt_thermal} shows that
$|G|^2 = \hbar \Im\chi_{F_\theta F_\theta}$. 
We thus have
\begin{equation}
\begin{split}
    S_{F_\theta F_\theta}^\t{sq}(\Omega)
    =\hbar\left[2N(\Omega)+1
    +2\mathrm{Re}\{M(\Omega)e^{-2i\theta}\}\right]
    \\
    \times
    \Im \chi_{F_\theta F_\theta}(\Omega).
\end{split}
\label{eq:squeezed_fdt_app}
\end{equation}
By comparison with the thermal FDT in \cref{eq:fdt_thermal}, we see that this 
squeezed-bath FDT can be expressed in terms of an effective occupation number
\begin{equation}
    n_\mathrm{eff}(\Omega)
    =N(\Omega)+\mathrm{Re}\{M(\Omega)e^{-2i\theta}\}.
\end{equation}
Thus, it is legitimate to interpret the effect of the squeezed bath through an effective
phonon occupation. 

\section{Equivalent phonon number estimation}

We estimate an equivalent phonon occupation of the feedback-defined mode by integrating the displacement power associated with physical motion near the trapped-mode resonance and referencing it to the zero-point motion of the confined oscillator.

\begin{figure*}[hbt!]
  \centering
  \includegraphics[width=0.95\textwidth]{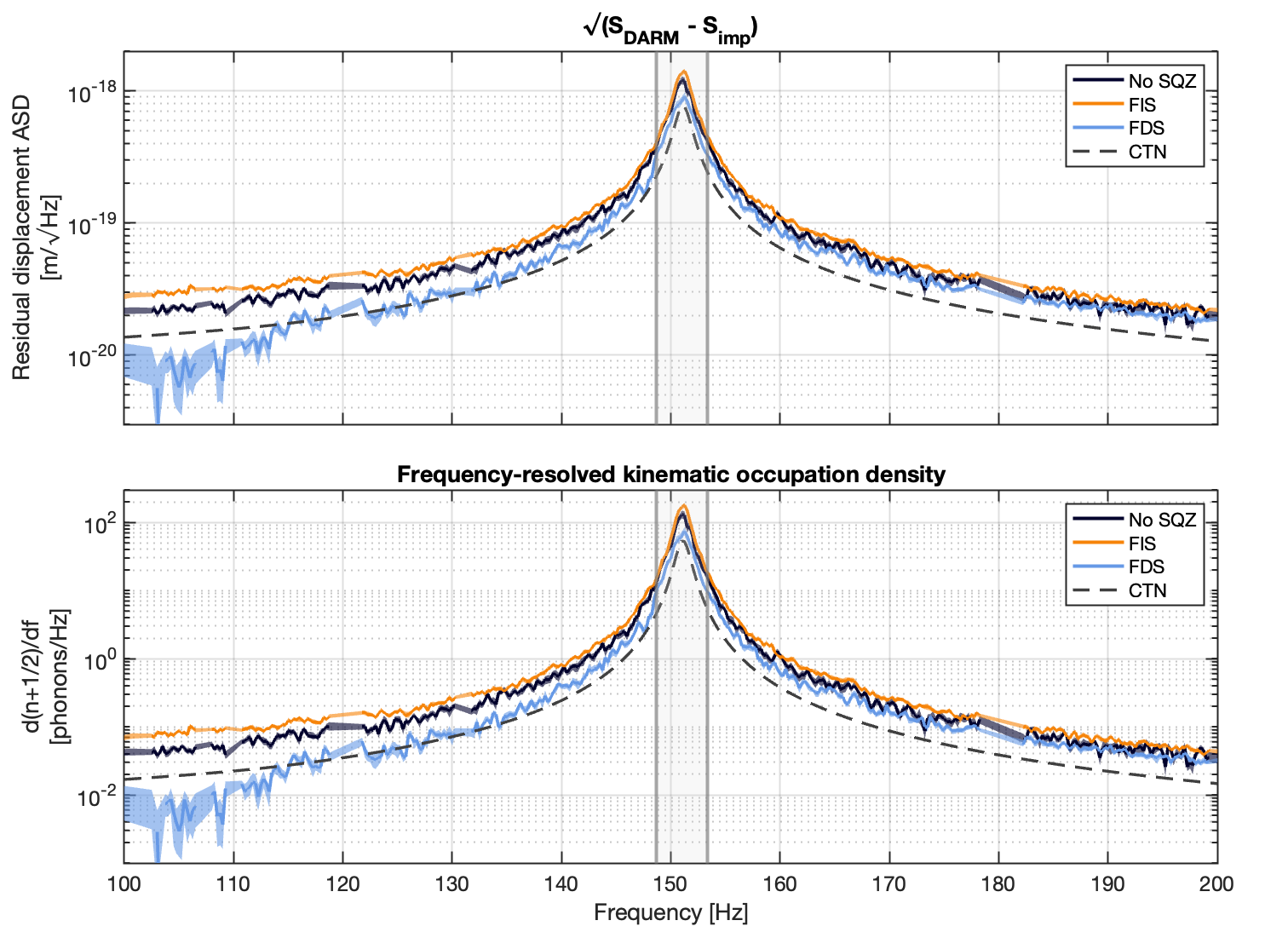}
  \caption{\textbf{Force equivalent ASD of the trapped mode.}The vertical grey lines indicate the 3 full-width half-max points around the resonance.}
  \label{fig:force_equiv}
\end{figure*}

\paragraph*{\textbf{Separating motion from imprecision.}}
To estimate the motion-only contribution, we subtract the output-referred imprecision component predicted by the quantum noise model. For each configuration, we compute the single-sided displacement PSD $S_{xx}^{\mathrm{meas}}(\Omega)$ from the calibrated DARM error signal and evaluate the corresponding imprecision PSD $S_{xx}^{\mathrm{imp}}(\Omega)$ from the model. We then define the inferred motion PSD as
\begin{equation}
S_{xx}^{\mathrm{mot}}(\Omega) \equiv \max\!\left[
S_{xx}^{\mathrm{meas}}(\Omega)
-
S_{xx}^{\mathrm{imp}}(\Omega),
\,0\right],
\end{equation}
with the subtraction performed in power (PSD), not at the ASD level. The subsequent force-equivalent spectra in ASD units are shown in Figure~\ref{fig:force_equiv} along with the corresponding error bars. 

\begin{figure*}[hbt!]
  \centering
  \includegraphics[width=\linewidth]{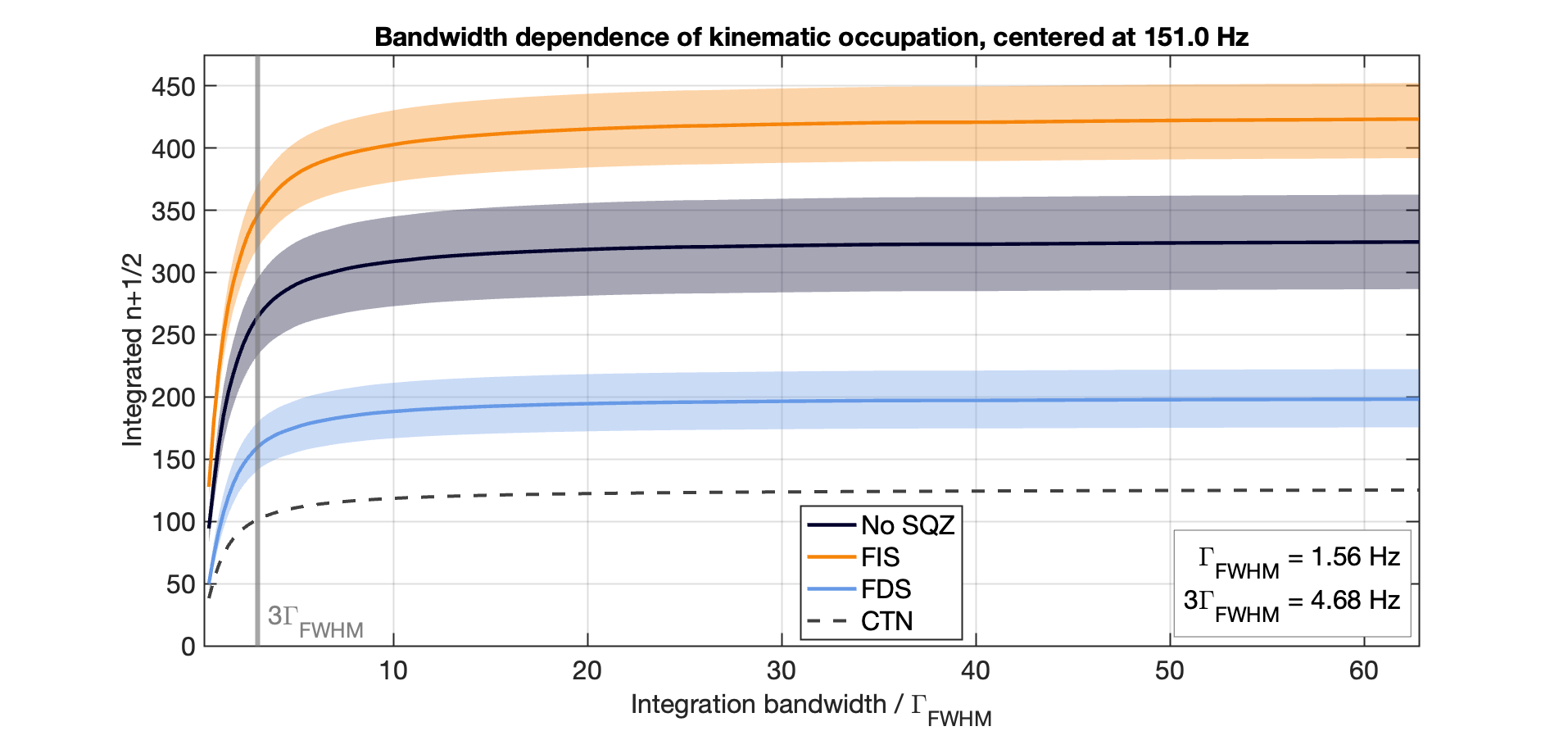}
  \caption{\textbf{Kinematic phonon occupation number as a function of integration bandwidth.} The vertical grey line indicates 3 full-width half-max width for the integration band.}
  \label{fig:kinematic_neff}
\end{figure*}

\paragraph*{\textbf{Integration around resonance.}}
We determine the resonance frequency $\Omega_\mathrm{eff}$ and linewidth $\Gamma_\mathrm{eff}$ of the trapped mode from a fit to the resonance peak. The mean-square displacement associated with motion in a narrow band around resonance is obtained by integrating the motion PSD,
\begin{equation}
\langle x_D^2\rangle_{\mathrm{mot}}
=
\int_{\Omega_\mathrm{eff}-60\Gamma_\mathrm{eff}}^{\Omega_\mathrm{eff}+60\Gamma_\mathrm{eff}}
S_{xx}^{\mathrm{mot}}(\Omega)\,\frac{d\Omega}{2\pi}.
\end{equation}
The equivalent phonon occupation is then
\begin{equation}
n_\mathrm{eff}
=
\frac{\langle x_D^2\rangle_{\mathrm{mot}}}{2x_\mathrm{zpf}^2}
-
\frac{1}{2},
\qquad
x_\mathrm{zpf}
=
\sqrt{\frac{\hbar}{2m\Omega_\mathrm{eff}}}.
\end{equation}
 
 Figure~\ref{fig:kinematic_neff} shows the effective phonon occupancy number calculated using the kinematic approximation, as a function of integration band-width. The integration bandwidth axis is expressed in terms of full-width half-max of the resonance. It is clear that the phonon estimation converges to a finite number with increasing integration bandwidth for each case. The corresponding error bars around the estimation are indicated with the colored bands around the solid lines. 

 Table~\ref{tab:neff_summary} shows the effective phonon occupancy number for all three input quantum states, choosing 60 full-width half-max as the integration bandwidth around the resonance. This value for the bandwidth is found appropriate due to the integration bandwidth vs estimated phonon number plot shown in Figure~\ref{fig:kinematic_neff}. This plot is obtained by integrating the area underneath the motion-only residual displacement shown in Figure~\ref{fig:force_equiv}.

\paragraph*{\textbf{Contribution of coating thermal noise (CTN) to equivalent phonon number.}}
In addition to quantum noise and the narrow trapped-mode motion, the calibrated displacement spectrum contains broadband displacement noise from classical sources, including coating thermal noise (CTN). Around the trapped-mode frequency ($\sim150$~Hz), CTN contributes an approximately smooth background beneath the resonance peak. We do not subtract CTN in the phonon-number estimate reported here. Instead, we treat CTN as a systematic contribution to the integrated displacement power and quote its expected magnitude based on the Advanced LIGO CTN estimate/noise budget. Because CTN is independent of the injected squeezing configuration, it does not account for the observed change between the FIS and FDS spectra concentrated at the trapped-mode resonance.

\begin{table}[hbt]
\centering
\caption{Equivalent phonon occupation estimated from the resonance-band displacement power after subtracting the model-predicted imprecision, along with the expected CTN contribution integrated over the same band. Quoted uncertainties represent the statistical uncertainty propagated from the measured displacement ASD (3.5\% in ASD, corresponding to 7\% in PSD).}

\label{tab:neff_summary}
\begin{tabular}{lcccc}
\hline\hline
Case & $f_0$ [Hz] & FWHM [Hz] & $n_\mathrm{eff}(\mathrm{meas-shot})$ & $n_\mathrm{eff}(\mathrm{CTN})$ \\
\hline
No SQZ & 151.081 & 1.499 & $324 \pm 38$ & $125 \pm 1.9$ \\
FIS    & 151.095 & 1.565 & $423 \pm 30$ & $125 \pm 1.9$ \\
FDS    & 151.108 & 1.551 & $198 \pm 23$ & $125 \pm 1.9$ \\ 
\hline\hline
\end{tabular}
\end{table}
\label{tab:phonons}

\paragraph*{\textbf{Statistical uncertainty.}}
We estimate the statistical uncertainty on $n_\mathrm{eff}$ from the measured ASD uncertainty of the calibrated displacement spectrum ($\sim$ 3.5\% in ASD for all configurations). Since the phonon estimate is proportional to an integral of the displacement \emph{power} spectral density, we propagate this as a 7\% fractional uncertainty in the integrated motion power, and quote the corresponding uncertainty on $n_\mathrm{eff}$ in Table~\ref{tab:neff_summary}.

\end{document}